\documentclass[preprintnumbers,article,amsmath,amssymb,floatfix,10pt,prd,onecolumn,
superscriptaddress,nofootinbib]{revtex4}
\usepackage[colorlinks=true, pdfstartview=FitV, linkcolor=blue, citecolor=red, urlcolor=magenta]{hyperref}
\usepackage{bbm}
\usepackage{amsfonts}
\usepackage{mathrsfs}
\usepackage{latexsym}
\usepackage{epsfig}
\usepackage{epstopdf}
\usepackage{epstopdf}
\usepackage{graphicx}
\usepackage{amssymb}
\usepackage{amsmath}
\usepackage{dcolumn}
\usepackage{bm}
\usepackage{color}
\usepackage{comment}
\usepackage{xcolor}
\begin{document}
\title{\bf Logarithm Corrections and Thermodynamics for Horndeski gravity like Black Holes}

\author{Riasat Ali}
\email{riasatyasin@gmail.com}
\affiliation{Department of Mathematics, Shanghai University, Shanghai-200444, Shanghai, People's Republic of China}

\author{Zunaira Akhtar}
\email{zunaira.math.pu@gmail.com}
\affiliation{Department of Mathematics, University of the Punjab, Quaid-e-Azam Campus, Lahore 54590, Pakistan}

\author{Rimsha Babar}
\email{rimsha.babar10@gmail.com}
\affiliation{Division of Science and Technology, University of Education, Township, Lahore-54590, Pakistan}

\author{G. Mustafa}
\email{gmustafa3828@gmail.com}
\affiliation{Department of Physics
Zhejiang Normal University
Jinhua 321004, People’s Republic of China}

\author{Xia Tiecheng}
\email{xiatc@shu.edu.cn}
\affiliation{Department of Mathematics, Shanghai University, Shanghai-200444, Shanghai, People's Republic of China}

\begin{abstract}
In this paper, we compute the Hawking temperature by applying quantum tunneling
approach for the Horndeski like black holes. We utilize the semi-classical phenomenon 
and WKB approximation to the Lagrangian field equation involving generalized 
uncertainty principle (GUP) and compute the tunneling rate as well as Hawking 
temperature. For the zero gravity parameter, we obtain results consistent without 
correction parameter or original tunneling. Moreover, we study the thermal 
fluctuations of the considered geometry and examine the stable state of the
system by heat capacity technique. We also investigate the behaviour of thermodynamic 
quantities under the influence of thermal fluctuations. We observe from the graphical 
analysis, the corresponding system is thermodynamically stable with these correction terms.\\\\
{\bf keywords:} Horndeski like black holes; Quantum gravity; Tunneling radiation, 
Thermal fluctuations, Corrected entropy, Phase transition.
\end{abstract}

\maketitle

\date{\today}

%%%%%%%%%%%%%%%%%%%%%%%%%%%%%%%%%%%%%%%%%%%%%%%%%%%%%%%%%%%%%%%%%%%%%%%%
%%%%%%%%%%%%%%%        Introduction        %%%%%%%%%%%%%%%%%%%%%%%%%%%%%
%%%%%%%%%%%%%%%%%%%%%%%%%%%%%%%%%%%%%%%%%%%%%%%%%%%%%%%%%%%%%%%%%%%%%%%%
\date{\today}
\maketitle

\section{Introduction}

Tunneling is the semi-classical mechanism in which particles have
radiated from black hole (BH) outer horizon. Some analysis
shows the keen interest in the Hawking temperature $(T_{H})$ via tunneling method from different BHs. The main aspect to examine the $T_{H}$ is the imaginary part of classical action which leads to the tunneling radiation of boson particles appearing from the Horndeski like BHs.

The quantum tunneling and $T_{H}$ of charged fermions in BHs has been observed \cite{1}. In this paper, they
examined that the tunneling and $T_{H}$ depend on charges of
electric and magnetic, acceleration, rotation, mass and NUT parameter of the charged pair BHs. The tunneling strategy
from Reissner Nordstom-de Sitter BH like solution in global monopole has been analyzed \cite{2}.
In this article, the authors observed that the modified $T_{H}$ depends
on the parameter of global monopole. The BH thermodynamics have been examined \cite{4} with some parameters like acceleration, NUT and rotation. The researchers studied thermodynamical
quantities like the area, entropy, surface gravity and $T_{H}$.
The tunneling spectrum of bosonic particles  has been computed from the modified
BHs horizon by utilizing the Proca field equation. Hawking evaluated tunneling probability from BH \cite{10} by utilizing theoretical technique and later, it has been explained by Parikh and Wilczek
\cite{11, 12}. The important of this radiation represents that vacuum thermal fluctuation
produce pairs of particle (particle and anti-particle) from the horizon. Hawking considered that the particle's
have ability to emit from the BH and the anti-particles have no ability to radiate from
the horizon. Parikh and Wilczek explained a mathematical approach by utilizing WKB approximation.
This phenomenon use geometrical optic approximation which is another view of eikonal approximation in
wave clarification \cite{13}. The set of all particles remain at the front boundary and with the emission of these particles, the BH mass reduces in the form of particles energy.

In the Parikh-Wilczek method, a precisely tunneling was established and there were as still
unanswered problems like information release, temperature unitary and divergence. Many authors
have made efforts on the tunneling strategy and semi-classical phenomenon from the different BHs horizon; one of the important explanations can be checked in \cite{15}-\cite{44}. The radiate particles for many
BHs have been analyzed and also computed the radiate particles with the influences of the geometry of BH with different parameters. It is possible to study modified thermodynamic
properties of BH by considering generalized uncertainty principle (GUP) influences \cite{45}.
The GUP implies high energy result to thermodynamic of BH, by considering the quantum gravity theory
with a minimal length. By considering the GUP influences, it is viable to examine
the modified thermodynamic of BHs.

It is a well known fact that thermal fluctuations are a result of statistical perturbations in the dense matter. With the emission of Hawking radiations from the BH, the size of BH reduces and consequently its temperature increases.  Faizal and his colleague \cite {[3]} have studied the thermodynamics and thermal fluctuations of  generalized Schwarschild BH, (i.e., Reissner-Nordstrom, Kerr and charged AdS BHs) with the help first-order corrections and discussed the stability of BHs. The thermodynamics of rotating Kerr-AdS BH and its phase transition have been studied by Pourhassan and Faizal \cite{[4]}. They concluded that the entropy corrections are very helpful to examine the geometry of small BHs. By applying the stability test of heat capacity and Hessian matrix, the phase transition as well as thermodynamics of non-minimal regular BHs in the presence of cosmological constant has been investigated \cite{[5]} and the authors concluded that the local and global stability of the corresponding BHs increases for higher values of correction parameters. Zhang and Pradhan \cite{[6], [7]} have investigated the corrected entropy and second order phase transition via thermal fluctuations on the charged accelerating BHs.

Moreover, the thermodynamics and geometrical analysis of new Schwarzschild BHs have been studied \cite{[13],[14]}.
By using the tunneling approach under the influence of quantum gravity the Hawking temperture for different types of BH have been discussed \cite{[14a]}-\cite{[14e]}. Sharif and Zunaira \cite{[15],[16]} have computed the thermodynamics, quasi-normal modeand thermal fluctuations of charged BHs with the help of Weyl corrections. The authors found that the system is unstable for the small radii of BHs under the influence of first order corrections and by using the heat capacity and Hessian matrix technique, they have also studied the stable conditions of the system. The authors in \cite{[17],[18]} have investigated the thermodynamics, phase transition and local/global stability of NUT BH via charged, accelerating as well as rotating pairs. Ilyas et al.\cite{[18a],[18b],[18c],[18d]} discussed the energy conditions and calculated the new solutions for stellar structures by taking black hole geometry as exterior spacetime in the background of different modified theories of gravity. Recently, Ditta et al. \cite{[18e]} discussed the thermal stability and Joule–Thomson
expansion of regular BTZ-like black hole.

The main intention of this paper is to investigate the tunneling radiation without self-gravity and
back-reaction and also explain the modified tunneling
rate. The tunneling radiation is evaluated under
the conditions of charge-energy conservation, Horndeski parameter and GUP parameter influences. The modified $T_{H}$ depends on the Horndeski parameter as well as GUP parameter and also investigated the behaviour of thermodynamic quantities via thermal fluctuations.

This paper is based on the analysis of quantum tunneling, $T_{H}$, stability and
instability conditions for the Horndeski like BH.  The paper is outlined as follows:
in Sec. \textbf{II}, we study the tunneling
radiation of bosonic particles for $4D$ Horndeski like BH and also calculate the effects of GUP parameters on tunneling and $T_{H}$. In section \textbf{III},
we study the graphical presentation of tunneling radiation for this type of BH
and analyze the stable and unstable conditions for Horndeski like BH. In section \textbf{IV}, we investigate the behaviour of thermodynamic quantities under the effects of thermal fluctuations.
In section \textbf{V}, we express the discussion and conclusion of the whole analysis.

\section{Horndeski Like Black Holes}

Hui and his coauthor Nicolis \cite{46a} argued that the no-hair theorems cannot be applied on a Galileon field, as it is coupled to gravity under the effect of peculiar derivative interactions.  Further, they demonstrated that static and spherically symmetric spacetime defining the geometry of the black hole could not sustain nontrivial Galileon profiles. Babichev and Charmousis \cite{46b} examined the no-hair theorem in Ref. \cite{46a} by considering Horndeski theories and beyond. Furthermore, they provided the Lagrangian of Horndeski theory which can be expressed as a generalized Galileon Lagrangian, which is defined as
\begin{eqnarray}
S=& \int \sqrt{-g}\left(Q_2(\chi)+Q_3(\chi) \square \phi+Q_4(\chi) R+Q_{4, \chi}\left[(\square \phi)^2\right]\right).\nonumber\\
&\left.-\left(\nabla^\epsilon \nabla^\varepsilon \phi\right)\left(\nabla_\epsilon \nabla_\varepsilon \phi\right)\right]+Q_5(\chi) G_{\epsilon \varepsilon} \nabla^\epsilon \nabla^\varepsilon \phi \nonumber\\
&-\frac{1}{6} Q_{5, \chi}\left[(\square \phi)^3-3(\square \phi)\left(\nabla^\epsilon \nabla^v \phi\right)\left(\nabla_\epsilon \nabla_\varepsilon \phi\right)\right.\nonumber\\
&\left.\left.+2\left(\nabla_\epsilon \nabla_\varepsilon \phi\right)\left(\nabla^v \nabla^\gamma \phi\right)\left(\nabla_\gamma \nabla^\epsilon \phi\right)\right]\right\} d^4 x .
\end{eqnarray}
where $Q_2$, $Q_3$, $Q_4$, and $Q_5$ are the arbitrary functions of the scalar field $\phi$ and $\chi=-\partial^\epsilon \phi \partial_\epsilon \phi / 2$ represents the canonical kinetic term. Additionally, in the current analysis, $f_\chi$ stands for $\partial f(\chi) / \partial \chi$, $G_{\epsilon \varepsilon}$ is the Einstein tensor$R$ is the Ricci scalar, and other relations are defined as:
\begin{eqnarray}
&\left(\nabla_\epsilon \nabla_\varepsilon \phi\right)^2 \equiv \nabla_\epsilon \nabla_\varepsilon \phi \nabla^\varepsilon \nabla^\epsilon \phi\nonumber \\
&\left(\nabla_\epsilon \nabla_\varepsilon \phi\right)^3 \equiv \nabla_\epsilon \nabla_\varepsilon \phi \nabla^\varepsilon \nabla^\rho \phi \nabla_\rho \nabla^\epsilon \phi
\end{eqnarray}
The scalar field admits the Galilean shift symmetry $\partial_\epsilon \phi \rightarrow$ $\partial_\epsilon \phi+b_\epsilon$ in flat spacetime for $Q_2 \sim Q_3 \sim \chi$ and $Q_4 \sim Q_5 \sim \chi^2$, which resembles the Galilean symmetry \cite{46c}. In the current study, we investigate the tunneling radiation of spin-$1$ massive boson particles from Horndeski-like BH. For this purpose, we adopted the procedure, which is already reported in \cite{46} for Horndeski spacetime. Finally, we have the following spacetime:
\begin{eqnarray}
ds^{2}&=&-\left(1-\frac{2r M(r)}{\Sigma}\right)dt^{2}
+\frac{1}{\nabla(r)}\Sigma dr^{2}+\Sigma^{2} d\theta^{2}-
\frac{A}{\Sigma}\sin^{2}d\phi^{2}
-\frac{4ar}{\Sigma} M(r)\sin^{2}\theta dtd\phi,\label{M}
\end{eqnarray}
with
$\Sigma^{2}=a^{2} \cos^{2}\theta+r^{2}$ and $\nabla(r)=a^{2}-2rM(r)+r^{2}$,
 $M(r)=M-\frac{1}{2}Q In{\frac{r}{r_{0}}}$ and $A=(a^{2}+r^{2})^{2}-\nabla a^{2}\sin^{2}\theta$, while $a$, $Q$
and $M$ represent the rotation parameter, Horndeski parameter and mass of BH,
respectively.
If $Q\rightarrow 0$, the metric (\ref{M}) goes over
to the Kerr BH \cite{47} and if $Q=a=0$ the metric (\ref{M}) also goes over
to the Schwarzschild metric. The line-element (\ref{M}) can be re-written as
\begin{equation}
ds^{2}=-f(r)dt^{2}+g^{-1}(r)dr^{2}+I(r)d\phi^{2}+
h(r) d\theta^{2}+2R(r)dtd\phi\label{M1}
\end{equation}
where
\begin{eqnarray*}
f(r)&=&\left(1-\frac{2rM(r)}{\Sigma}\right),
~~~g{-1}(r)=\frac{1}{\nabla}\Sigma,~~~
h(r)=\Sigma^{2},\nonumber\\I(r)&=&-
\frac{A}{\Sigma}\sin^{2},~~~
R(r)=-\frac{2ar}{\Sigma} M(r)\sin^{2}\theta.
\end{eqnarray*}
We study the tunneling radiation of spin-1 particles from four-dimensional Horndeski like BHs. By utilizing the Hamilton-Jacobi ansatz and the WKB approximation to the
modified field equation for the Horndeski space-time, the tunneling phenomenon
is successfully applied. We study the modified filed equation on a four dimensional
space-time with the background of rotation parameter, Horndeski parameter and evaluated for the
radial function. As a result, we get the tunneling probability of the radiated
particles and derive the modified $T_{H}$ of Horndeski like BHs.
The modified filed equation is expressed by \cite{39, 43}
\begin{eqnarray}
&&\partial_{\mu}\Big(\sqrt{-g}\Psi^{\nu\mu}\Big)+\sqrt{-g}\frac{m^2}{\hbar^2}\Psi^{\nu}+\sqrt{-g}\frac{i}{\hbar}A_{\mu}\Psi^{\nu\mu}
+\sqrt{-g}\frac{i}{\hbar}eF^{\nu\mu}\Psi_{\mu}+\hbar^{2}\beta\partial_{0}\partial_{0}\partial_{0}\Big(\sqrt{-g}g^{00}\Psi^{0\nu}\Big)\nonumber\\
&&-\hbar^{2}\beta \partial_{i}\partial_{i}\partial_{i}\Big(\sqrt{-g}g^{ii}\Psi^{i\nu}\Big)=0,\label{bbb}
\end{eqnarray}
here $\Psi^{\nu\mu}$, $m$ and $g$ present the anti-symmetric tensor, bosonic particle mass and determinant of coefficient matrix, so
\begin{eqnarray}
\Psi_{\nu\mu}&=&\Big(1-\hbar^2\beta\partial_{\nu}^2\Big)\partial_{\nu}\Psi_{\mu}-\Big(1-\hbar^2\beta\partial_{\mu}^2\Big)\partial_{\mu}\Psi_{\nu}
+\Big(1-\hbar^2\beta\partial_{\nu}^2\Big)\frac{i}{\hbar}eA_{\nu}\Psi_{\mu}\nonumber\\
&-&\Big(1-\hbar^2\beta\partial_{\nu}^2\Big)\frac{i}{\hbar}eA_{\mu}\Psi_{\nu},~~\textmd{and}~~
F_{\nu\mu}=\nabla_{\nu} A_{\mu}-\nabla_{\mu} A_{\nu},\nonumber
\end{eqnarray}
with $\beta$, $e$~, $\nabla_{\mu}$ and $A_{\mu}$ are the GUP parameter(quantum gravity), bosonic particle
charge, covariant derivative and BH potential, respectively.
The $\Psi^{\nu\mu}$ can be computed as
\begin{eqnarray}
\Psi^{0}&=&\frac{-I\Psi_{0}+R\Psi_{3}}{fI
+R^2},~~~~~\Psi^{1}=\frac{1}{g^{-1}}\Psi_{1},~~~\Psi^{2}
=\frac{1}{h}\Psi_{2},~~~\Psi^{3}=\frac{R\Psi_{0}
+f\Psi_{3}}{fI+R^2},~~~~~\Psi^{01}
=\frac{\tilde{-D}\Psi_{01}+R\Psi_{13}}{(R^2+fI
)g^{-1}},\nonumber\\~~~\Psi^{02}&=&\frac{\tilde{-D}\Psi_{02}}{
(R^2+fI)h},~~~~\Psi^{03}=\frac{(f^2-fI)\Psi_{03}}{(fI+R^2)^2},~~\Psi^{12}=\frac{1}{g^{-1}
h}\Psi_{12},~~\nonumber\\
\Psi^{13}&=&\frac{1}{g^{-1}(f
I+R^2)}\Psi_{13},~~\Psi^{23}=\frac{f\Psi_{23}+R\Psi_{02}}
{(fI+R^2)h}.\nonumber
\end{eqnarray}
In order to observe the bosonic tunneling, we have assumed Lagrangian gravity equation. Further, we utilized the WKB
approximation to the Lagrangian gravity equation and computed set of equations. Furthermore, we have utilized the variable separation action to get required solutions. The approximation of WKB is defined \cite{5} as
\begin{equation}
\Psi_{\nu}=\eta_{\nu}\exp\left[\frac{i}{\hbar}K_{0}(t,r,\phi, \theta)+
\Sigma \hbar^{n}K_{n}(t,r, \phi, \theta)\right].\label{ab}
\end{equation}
we get set of equations in \textbf{Appendix A}.
Utilizing variable separation technique, we can take
\begin{equation}
K_{0}=-(E-L\omega)t+W(r)+L\phi+\nu(\theta),\label{ff}
\end{equation}
where $E$ and $L$ present the particle energy and particle angular, respectively, corresponding to angle $\phi$.

After considering Eq. (\ref{ff}) into Eqs. (\ref{j1})-(\ref{j2}), we reach a matrix in the form
\begin{equation*}
U(\eta_{0},\eta_{1},\eta_{2},\eta_{3})^{T}=0,
\end{equation*}
which express a $4\times4$ matrix presented as "$U$", whose elements
are given as follows:
\begin{eqnarray}
U_{00}&=&\frac{-I}{g^{-1}(fI+R^2)}
\left[W_{1}^2+\beta W_{1}^4\right]-\frac{I}{(fI+R^2)h}\Big[L^2+\beta L^4\Big]
-\frac{fI}{(fI+R^2)^2}
\Big[\nu_{1}^2+\beta \nu_{1}^4\Big]-\frac{m^2 I}{(f
I+R^2)},\nonumber\\
U_{01}&=&\frac{-I}{g^{-1}(fI+R^2)}
\Big[((E-L\omega)+ (E-L\omega)^3\beta+A_{0}e+(E-L\omega)^2\beta eA_{0}\Big]W_{1}
+\frac{R}{g^{-1}(fI+R^2)}+\Big[\nu_{1}+
\beta \nu_{1}^3\Big],\nonumber\\
U_{02}&=&\frac{-I}{h(fI+R^2)}
\Big[(E-L\omega)+ (E-L\omega)^3\beta-A_{0}e-(E-L\omega)^2\beta eA_{0}
\Big]L,\nonumber\\
U_{03}&=&\frac{-R}{g^{-1}(fI+R^2)}
\Big[W_{1}^2+\beta W_{1}^4\Big]- \frac{fI}{h
(fI+R^2)^2}\Big[(E-L\omega)^3\beta
-(E-L\omega)^2\beta eA_{0}+(E-L\omega)-eA_{0}\Big]\nu_{1}\nonumber\\&+&\frac{m^2R}{(fI+R^2)^2},\nonumber\\
U_{12}&=&\frac{1}{g^{-1}h}\Big[W_{1}+\beta W_{1}^3\Big]L,\nonumber\\
U_{11}&=&\frac{-I}{g^{-1}(fI+R^2)}
\Big[\beta (E-L\omega)^4-\beta eA_{0}EW_{1}^2+(E-L\omega)^2-eA_{0}(E-L\omega)\Big]
+\frac{R}{(fI+R^2)g^{-1}}\nonumber\\
&+&\Big[\nu_{1}+\beta \nu_{1}^3\Big](E-L\omega)-\frac{1}{g^{-1}
h}\Big[L^2+\beta L^4\Big]-\frac{1}{(fI
+R^2)g^{-1}}\Big[\nu_{1}+\beta \nu_{1}^3\Big]-\frac{m^2}{g^{-1}}
-\frac{eA_{0}I}{(fI+R^2)g^{-1}}\nonumber\\
&\times&
\Big[(E-L\omega)+ (E-L\omega)^3\beta-A_{0}e-(E-L\omega)^2\beta eA_{0}\Big]
+\frac{eA_{0}R}{g^{-1}(fI+R^2)}\Big[\nu_{1}+
\beta \nu_{1}^3\Big],\nonumber\\
U_{13}&=&\frac{-R}{g^{-1}(fI+R^2)}\Big[W_{1}
+\beta W_{1}^3\Big](E-L\omega)+\frac{1}{g^{-1}(fI
+R^2)^2}\Big[W_{1}+\beta W_{1}^3\Big]\nu_{1}+\frac{ReA_{0}}
{g^{-1}(fI+R^2)}\Big[W_{1}+\beta W_{1}^3\Big],\nonumber\\
U_{20}&=&\frac{I}{h(fI+R^2)}
\Big[(E-L\omega)L+\beta (E-L\omega)L^3\Big]+\frac{R}{h
(fI+R^2)}\Big[(E-L\omega)+\beta (E-L\omega)^3\nu_{1}^2\Big]
\nonumber\\&-&\frac{IeA_{0}}{h(fI+R^2)}\Big[L+\beta L^3\Big],\nonumber\\
U_{22}&=&\frac{I}{h(R^2+fI)}\Big[\beta E^4-\beta eA_{0}E+E^2
-eA_{0}(E-L\omega)\Big]
-\frac{1}{g^{-1}h}+\frac{R}{h(R^2+fI)}\Big[(E-L\omega)^3\beta\nonumber\\
&+&-(E-L\omega)^2\beta eA_{0}-A_{0}e+(E-L\omega)\Big]\nu_{1}
-\frac{f}{h(R^2+fI)}\Big[\nu_{1}^2+
\beta \nu_{1}^4\Big]-\frac{m^2}{h}-\frac{eA_{0}I}{h
(fI+R^2)}
\nonumber\\&&
\Big[(E-L\omega)+ (E-L\omega)^3\beta-A_{0}e-(E-L\omega)^2\beta eA_{0}\Big],\nonumber\\
U_{23}&=&\frac{f}{h(fI+R^2)}
\Big[L+\beta L^3\Big]\nu_{1},\nonumber\\
U_{30}&=&\frac{(fI-f^2)}{(fI
+R^2)^2}\Big[\nu_{1}+\beta \nu_{1}^3\Big]E
+\frac{R}{h(fI+R^2)}
\Big[L^2+\beta L^4\Big]-\frac{m^2R}{(fI
+R^2)}-\frac{eA_{0}(fI-f^2)}{(fI
+R^2)^2}\Big[\nu_{1}+\beta \nu_{1}^3\Big],\nonumber\\
U_{31}&=&\frac{1}{g^{-1}(fI+R^2)}
\Big[\nu_{1}+\beta \nu_{1}^3\Big]W_{1},\nonumber\\
U_{32}&=&\frac{R}
{h(R^2+fI)}\Big[L+\beta L^3\Big]
E+\frac{f}{h(R^2+fI)}
\Big[\nu_{1}+\beta \nu_{1}^3\Big]L,\nonumber\\
U_{33}&=&\frac{(fI-f^2)}{(fI
+R^2)}\Big[(E-L\omega)^2-eA_{0}(E-L\omega)
+\beta (E-L\omega)^4-\beta eA_{0}(E-L\omega)^3\Big]\nonumber\\&-&\frac{1}{g^{-1}(R^2+fI
)}\Big[W_{1}^2+\beta W_{1}^4\Big]
-\frac{f}{(R^2+fI)h}
\Big[L^2+\beta L^4\Big]-\frac{m^2 f}{(fI+R^2)}
-\frac{eA_{0}(fI-f^2)}{(fI
+R^2)}\nonumber\\&\times&\Big[(E-L\omega)+\beta (E-L\omega)^3-eA_{0}(E-L\omega)^2\Big],\nonumber
\end{eqnarray}
with $\partial_{t}K_{0}=(E-L\omega),~~\partial_{\phi}K_{0}=L$,
$W_{1}=\partial_{r}{K_{0}}$ and $\nu_{1}=\partial_{\theta}{K_{0}}$.
For non-trivial solution, we get
\begin{eqnarray}\label{a1}
ImW^{\pm}&=&\pm \int\sqrt{\frac{\Big(E-L\Omega-eA_{0}\Big)^{2}
+Z_{1}\left[1+\beta\frac{Z_{2}}{Z_{1}}\right]}{(fI+R^{2})gI^{-1}}}dr,\nonumber\\
&=&\pm i\pi\frac{\Big(R-L\Omega-A_{0}e\Big)+\Big[1+\beta A\Big]}{2k(r_{+})},
\end{eqnarray}
where
\begin{eqnarray}
Z_{1}&=&(E-L\omega)\nu_{1}\frac{g^{-1}R}{fI+R^{2}}+\frac{fg^{-1}}{fI+R^{2}}\nu^{2}_{1}-g^{-1}m^2,\nonumber\\
Z_{2}&=&\frac{g^{-1}I}{fI+R^{2}}\big[(R-L\Omega)^{4}+(eA_{0})^{2}(R-L\Omega)^{2}-2A_{0}e(R-L\Omega)^{3}\big]\nonumber\\
&+&\frac{g^{-1}R}{h(fI+R^{2})}\big[(R-L\Omega)^{3}-eA_{0}(R-L\Omega)^{2}\big]\nu_{1}-\frac{fg^{-1}}{fI+R^{2}}\nu^{4}_{1}-W^{4}_{1}.\nonumber
\end{eqnarray}
and $A$ is a arbitrary parameter. In particular case, we take the radial component of the action of particle, for this aim we choose a components of matrix equals to zero. Since, we have found the tunneling radiation (related to Horndeski gravity and quantum gravity) for BH.
This tunneling and $T_{H}$ quantities relate on the Horndeski gravity and quantum gravity of this particular physical object. Thus, we have found the corresponding $T_{H}$ which important as a component of leading metric with Horndeski gravity and quantum gravity. Such that, in this method we are not concerned in
order of higher for Planck's constant only obtained appropriate result. The generalized tunneling depends on the BHs metric and Horndeski gravity and GUP parameter.
The generalized tunneling for Horndeski like BH can be written as
\begin{equation}
T=\frac{T_{emission}}{T_{absorption}}=
\exp\left[{-2\pi}\frac{(R-L\Omega-A_{0}e)}
{k(r_{+})}\right]\left[1+\beta A\right],
\end{equation}
with
\begin{equation}
k(r_{+})=\frac{ 4\pi  r_+ \left(a^2+r_+^2\right)}{Q r_++-a^2+r_+^2}
\end{equation}
In the presence of GUP terms, we calculate the
$T_{H}$ of the Horndeski gravity BHs.
by taking the Boltzmann factor
$T_{B}=\exp\left[(E-L\omega-eA_{0})/T_{H}\right]$ as
\begin{eqnarray}
T_{H}=\frac{-a^2+Q r_{+}+r_+^2}{4 \pi  r_+ \left(a^2+r_{+}^{2}\right)}\left[1-\beta A\right].\label{b1}
\end{eqnarray}
The above result shows that the $T_{H}$
depends on the Horndeski gravity, GUP parameter,
rotation parameter, Horndeski gravity, arbitrary parameter $A$ and radius ($r_+$) of BH. When $\beta=0$, we obtain the general $T_{H}$ in \cite{46}. In the absence of charge i.e., $Q=0$, the above temperature reduces into Kerr BH temperature \cite{46a, 46b}. For $\beta=0$ and $a=0$, the temperature reduces into Reissner Nordstr$\ddot{o}$m BH. Moreover, when $Q=0=a$, we recover the temperature of Schwarzschild BH \cite{46c}. The quantum corrections slow down the increase in $T_{H}$ throughout the radiation phenomenon.

\subsection{$T_{H}$ versus $r_{+}$}

We observe the geometrical presentation of $T_{H}$
w.r.t $r_{+}$ for the $4D$ Horndeski like metric. Moreover, we observe the physical significance of these graphs under Horndeski gravity and GUP parameter and study the stability and instability analysis of corresponding  $T_{H}$. For $\beta$ equals to zero, the tunneling radiation will be independent of GUP parameter.
In the left plot of \textbf{Fig. 1}, the $T_{H}$ increases with increasing $\beta$ in  small region of horizon $0\leq r_{+}\leq 5$, that indicates the stable state of BH till $r_{+}\rightarrow \infty$.
In the right plot of \textbf{Fig. 1}, the rotating parameter and $\beta$ are fixed, then we take changing values of hairy parameter of Horndeski gravity and get the completely unstable form of BH with negative temperature.

\begin{center}
\includegraphics[width=7cm]{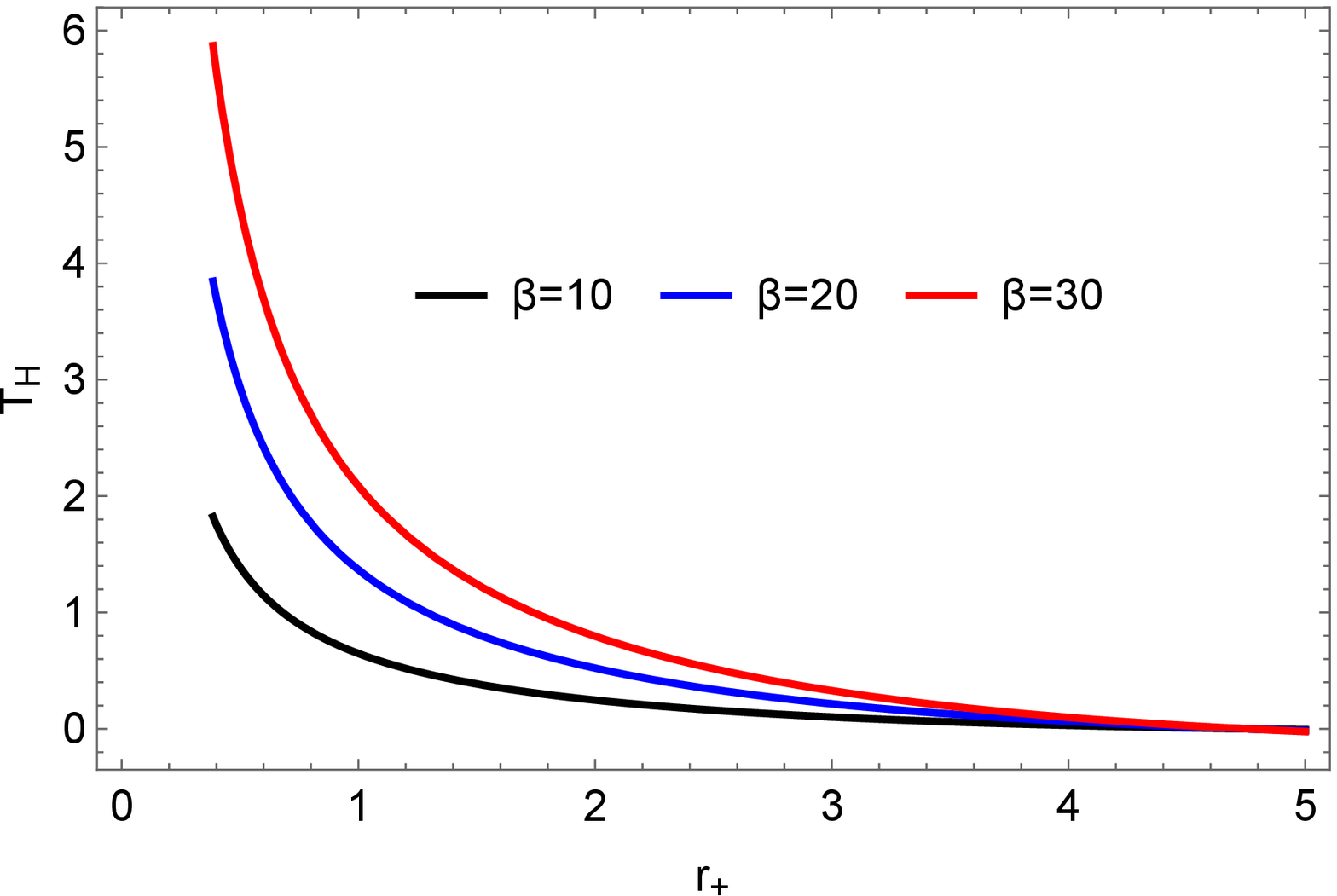}\includegraphics[width=7cm]{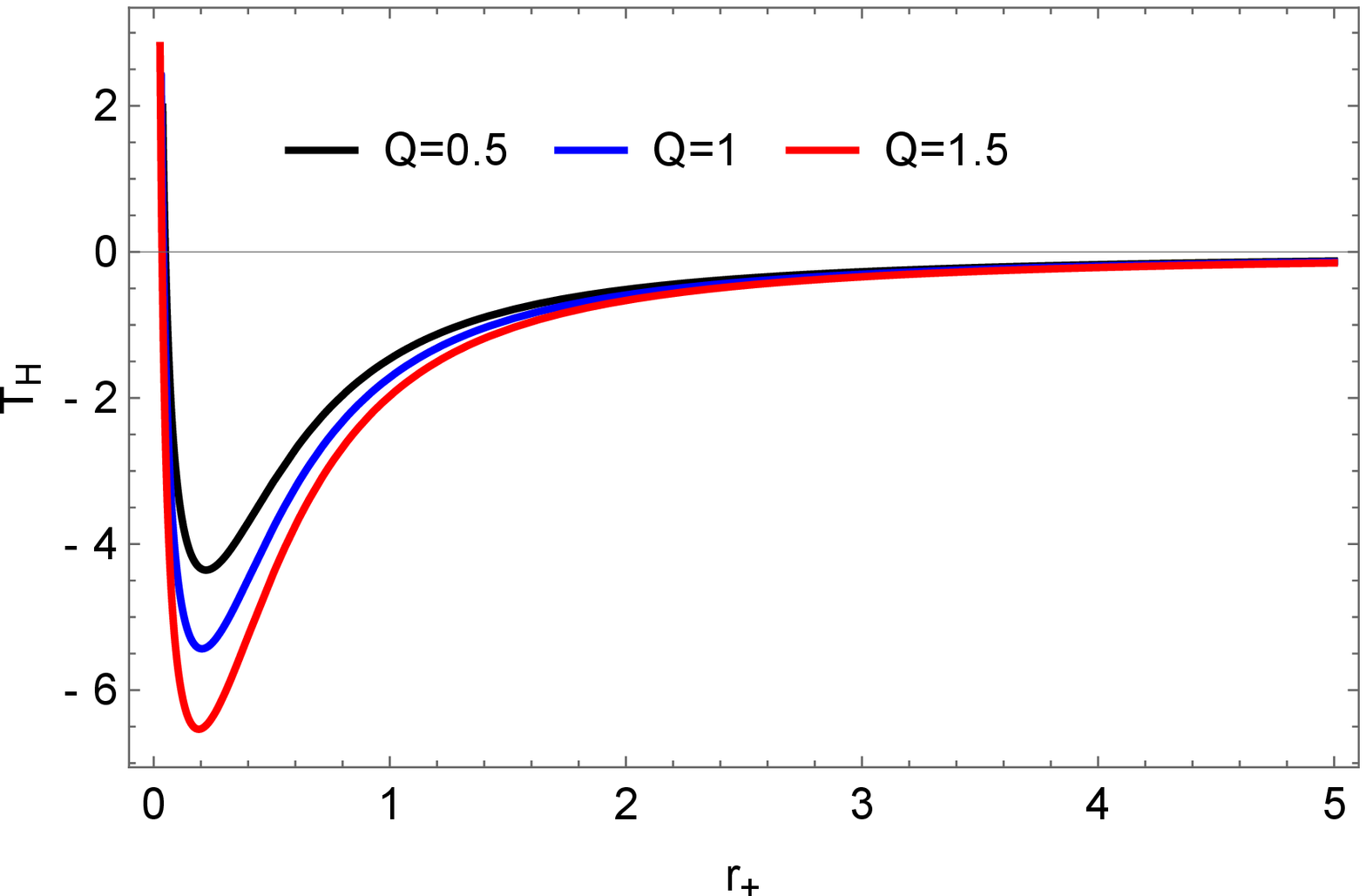}\\
{Figure 1: $T_{H}$ w.r.t horizon $r_{+}$ for $a=5$, $Q=0.5$ and $\Xi=1$ and left $\beta=10$ (black), $\beta=20$ (blue), $\beta=30$ (red). Right $a=0.5$, $\Xi=1$, $\beta=5$, $Q=0.5$ (black), $a=0.5$, $\Xi=1$, $\beta=5$, $Q=1$ (blue), $a=0.5$, $\Xi=1$, $\beta=5$, $Q=1.5$ (red).}
\end{center}

\section{Thermodynamics and Effects of first order corrections}
Thermal fluctuations plays important role on the study of BH thermodynamics. With the concept of Euclidean quantum gravity, the temporal coordinates shifts towards complex plan. To check the effects of these correction in entropy, we find Hawking temperature and usual entropy of the given system with the help of first law of thermodynamics
\begin{eqnarray}\label{2}
S=\pi  \left(a^2+r_+^2\right), \quad
T=\frac{-a^2+Q r_++r_+^2}{4 \pi  r_+ \left(a^2+r_+^2\right)}
\end{eqnarray}
 To check the corrected entropy along these thermal fluctuations, the partition function is $Z(\mu)$ in terms of  density of states $\eta(E)$ is given as \cite{[7]}
\begin{equation}
Z(\mu)=\int_{0}^{\infty} \exp(-\mu E)\eta(E)dE,
\end{equation}
where $T_{+}=\frac{1}{\mu}$ and E is the mean energy of thermal radiations. By using the Laplace inverse transform, the expression of density takes the form
\begin{equation}
\rho(E)=\frac{1}{2\pi
i}\int_{\mu_{0}-i\infty}^{\mu_{0}+i\infty} Z(\mu)
\exp(\mu E) d\mu=\frac{1}{2\pi
i}\int_{\mu_{0}-i\infty}^{\mu_{0}+i\infty}
\exp(\tilde{S}(\mu))d\mu,
\end{equation}
where $\tilde{S}(\mu)=\mu E+ \ln Z(\mu)$ represents the modified entropy of the considered system that is dependent on Hawking temperature. Moreover, the expression of entropy gets modified with the help of steepest decent method,
\begin{equation}
\tilde{S}(\mu)=S+\frac{1}{2}(\mu-\mu_{0})^{2}
\frac{\partial^{2}\tilde{S}(\mu)}{\partial
\mu^{2}}\Big|_{\mu=\mu_{0}}+\text{higher-order terms}.
\end{equation}
Using the conditions $\frac{\partial
\tilde{S}}{\partial\mu}=0$ and $\frac{\partial^{2}
\tilde{S}}{\partial\mu^{2}}>0$, the corrected entropy relation under the first-order corrections modified. By neglecting higher order terms, the exact expression of entropy is expressed as
\begin{equation}\label{4}
\tilde{S}=S-\delta \ln(ST^{2}),
\end{equation}
where $\delta$ is called correction parameter, the usual entropy of considered system is attained by fixing $\delta=0$ that is without influence of these corrections. Furthermore, inserting the Eq. (\ref{2})  into (\ref{4}), we have
\begin{eqnarray}
\tilde{S}&=&(a^2+r_+^2)-\delta  \log \left(\frac{\left(a^2-r_+ \left(Q+r_+\right)\right)^2}{16 \pi  r_+^2
   \left(a^2+r_+^2\right)}\right).
\end{eqnarray}
\begin{center}
\includegraphics[width=8cm]{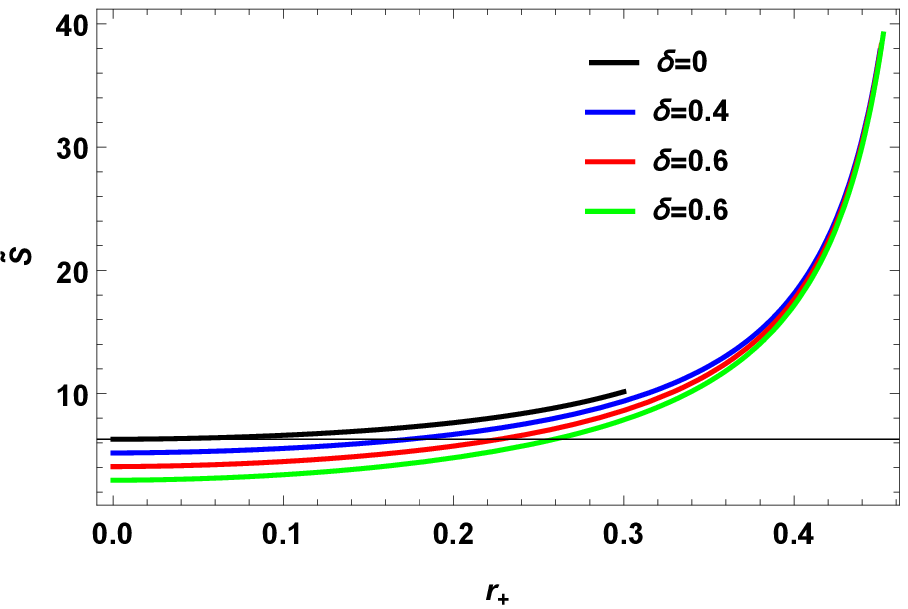}\\
{Figure 2: Corrected entropy versus $r_{+}$ for a=0.2, Q=0.4. }.
\\
\end{center}

In the \textbf{Fig. 2}, the graph of corrected entropy is monotonically increasing throughout the considered domain. It is noted the graph (black) of usual entropy is increasing just for small value of horizon radius but corrected expression of energy is increasing smoothly. Thus, these corrections terms are more effective for small BHs. Now, using the expression of corrected entropy and check the other
 other thermodynamic quantities via thermal fluctuations. In this way, the the Helmholtz energy ($F=-\int \tilde{S}dT$) leads to the form
\begin{eqnarray}
F&=&\frac{\left(a^4-r_+^2 \left(-4 a^{2}+2 Q r_{+}+r_+^2\right)\right) \left(\delta  \log \left(\frac{\left(a^2-r_+
   \left(Q+r_+\right)\right)^2}\{r_+^2 \left(a^2+r_+^2\right)\}\right)-a^2-\delta  \log (16 \pi
   )-r_+^2\right)}{4 \pi  r_+^2 \left(a^2+r_+^2\right)^2}.
\end{eqnarray}
 \begin{center}
\includegraphics[width=8cm]{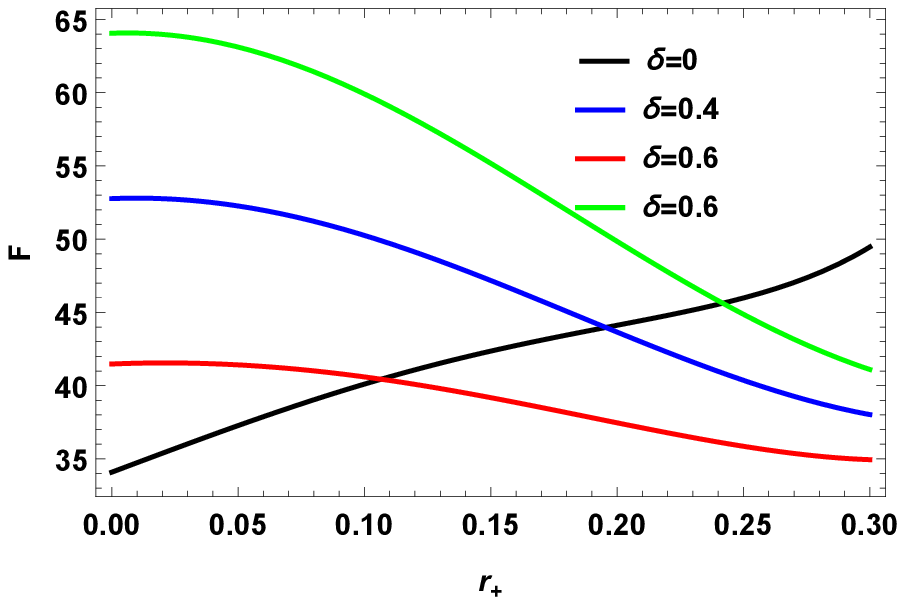}\\
{Figure 3: Helmholtz free energy versus $r_{+}$ for a=0.2, Q=0.4.}.
\\
\end{center}
The \textbf{Fig. 3} shows the graph of Helmholtz free energy versus horizon radius. It is observed that the behaviour of energy is gradually decreases for the different correction parameter $\delta$ values. While the graph of usual entropy shows opposite behaviour as the graph is increasing. This behaviour means, the considered system shifts its state towards equilibrium, thus, no more work can be extract from it. The expression of internal energy ($E=F+T\tilde{S}$) for the corresponding geometry is given by \cite{[7]}
\begin{eqnarray}
E&=&\Big(r_+ \left(a^2+r_+^2\right) \left(r_+ \left(Q+r_+\right)-a^2\right) \left(\delta  \left(\log (16 \pi
   )-\log \left(\frac{\left(a^2-r_+ \left(Q+r_+\right)\right){}^2}{r_+^2
   \left(a^2+r_+^2\right)}\right)\right)+\pi  \left(a^2+r_+^2\right)\right)\nonumber\\&-&\left(a^4-r_+^2 \left(-4 a^2+2 Q
   r_++r_+^2\right)\right) \left(-\delta  \log \left(\frac{\left(a^2-r_+
   \left(Q+r_+\right)\right)^2}{r_+^2 \left(a^2+r_+^2\right)}\right)+a^2+\delta  \log (16 \pi
   )+r_+^2\right)\Big)\nonumber\\&&\Big(4 \pi  r_+^2 \left(a^2+r_+^2\right)^2\Big)^{-1}.
\end{eqnarray}
\begin{center}
\includegraphics[width=8cm]{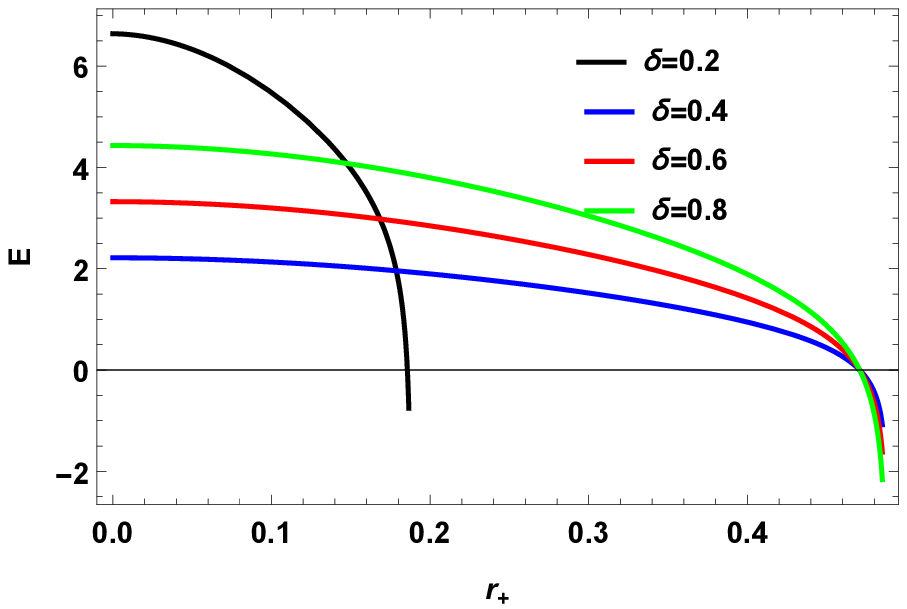}\\
{Figure 4: Internal Energy w.r.t $r_{+}$ for a=0.2, Q=0.4.}
\\
\end{center}
The graphical behaviour of internal energy for the different choices of horizon radius is shown in \textbf{Fig. 4}. It is observable that for the small values of radii, the graph is gradually decreases even shifts towards negative side, While the corrected internal energy depicts positive behaviour. This mean that the considered BH absorbing more and more heat from the surrounding to maintain its state. Since, BHs considered as a thermodynamic system, so there is another important thermodynamic quantity that is pressure. In this regard, there is deep connection between voulme ($V=\frac{2\pi(r_{+}^{2}+a^{2})(2r_{+}^{2}+a^{2})}{3r_{+}}$) and pressure. The Expression of BH pressure ($P=-\frac{dF}{dV}$) under the effect of thermal fluctuations takes the form
\begin{eqnarray}
P&=&\Big(2 r_+^2 \Big(a^2+r_+^2\Big) \Big(-4 a^2+3 Q r_++2 r_+^2\Big) \Big(a^2-r_+
   \Big(Q+r_+\Big)\Big)^2 \Big(-\delta  \log \Big(\Big(\Big(a^2-r_+
   (Q+r_+)\Big)^2\Big)(r_+^2 (a^2+r_+^2))^{-1}\Big)\nonumber\\&+&a^2+\delta  \log (16 \pi
   )+r_+^2\Big)-2 \Big(r_+ \Big(Q+r_+\Big)-a^2\Big) \Big(a^4-r_+^2 \Big(-4 a^2+2 Q
   r_++r_+^2\Big)\Big) \Big(-a^4 \delta +Q r_+^3 \Big(a^2+\delta \Big)\nonumber\\&-&r_+^2 \Big(a^4+3 a^2 \delta
   \Big)+Q r_+^5+r_+^6\Big)+4 r_+^2 \Big(a^4-r_+^2 \Big(-4 a^2+2 Q r_++r_+^2\Big)\Big)
   \Big(a^2-r_+ \Big(Q+r_+\Big)\Big)^2 \Big(-\delta  \log \nonumber\\&\times&\Big(\Big(\Big(a^2-r_+
   \Big(Q+r_+\Big)\Big)^2\Big)\Big(r_+^2 \Big(a^2+r_+^2\Big)\Big)^{-1}\Big)+a^2+\delta  \log (16 \pi
   )+r_+^2\Big)+2 \Big(a^2+r_+^2\Big) \Big(a^4-r_+^2 \Big(-4 a^2
   \nonumber\\&+&2 Q r_{+}+r_+^2\Big)\Big)\Big(a^2-r_+ \Big(Q+r_+\Big)\Big)^2 \Big(-\delta  \log \Big(\Big(\Big(a^2-r_+
   \Big(Q+r_+\Big)\Big)^2\Big)(r_+^2 \Big(a^2+r_+^2\Big)\Big)^{-1}\Big)+a^2+\delta  \log (16 \pi)\nonumber\\&+&r_+^2\Big)\Big) \Big(4 \pi  r_+^3\Big(a^2+r_+^2\Big)^3 \Big(a^2-r_+ \Big(Q+r_+\Big)\Big)^2\Big)^{-1}.
   \end{eqnarray}
 \begin{center}
\includegraphics[width=8cm]{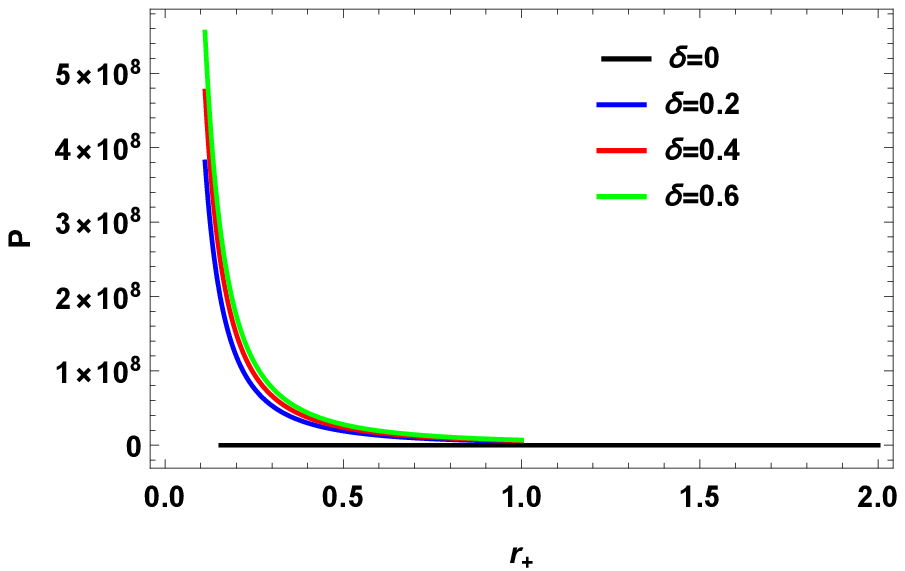}\\
{Figure 5: Pressure versus $r_{+}$ for a=0.2, Q=0.4.}
\end{center}
In the \textbf{Fig. 5}, the graph of pressure is just coincides the state of equilibrium. For the different values of correction parameter, the pressure is significantly increases for the considered system. Further, there is another important thermodynamic quantity enthalpy ($H=E+PV$) is given in \textbf{Appendix B}.
   \begin{center}
\includegraphics[width=8cm]{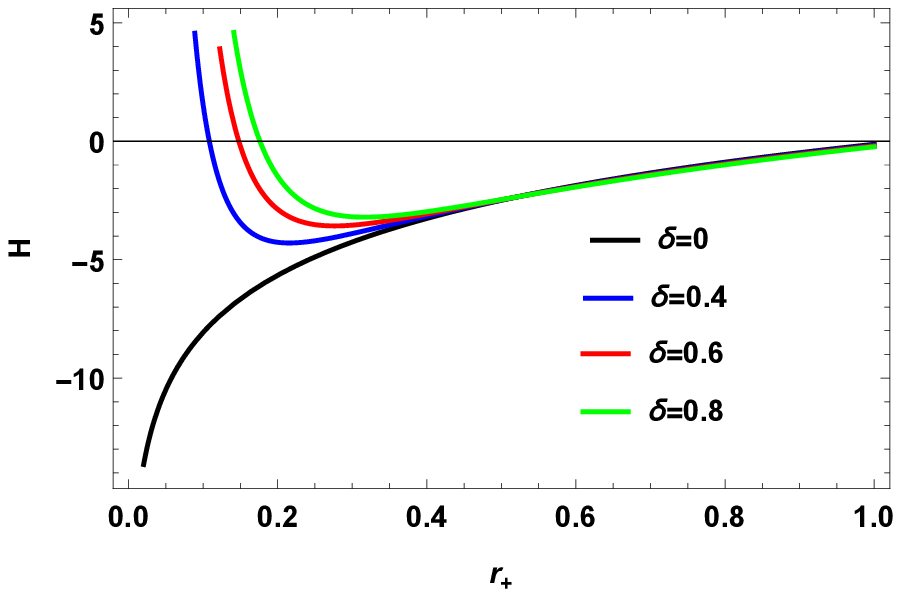}\\
{Figure 6: Enthalpy versus $r_{+}$ for a=0.2, Q=0.4.}
\end{center}
From \textbf{Fig. 6}, it can observed that the graph of usual enthalpy is coincide with the plots of corrected one and abruptly decreases even shifts towards negative side. This means that there exists a exothermic reactions means there will be huge amount of energy release into its surroundings. By taking into account the thermal fluctuations, the Gibbs free energy ($G=H-T\tilde{S}$) is expressed in \textbf{Appendix B}.
\begin{center}
\includegraphics[width=8cm]{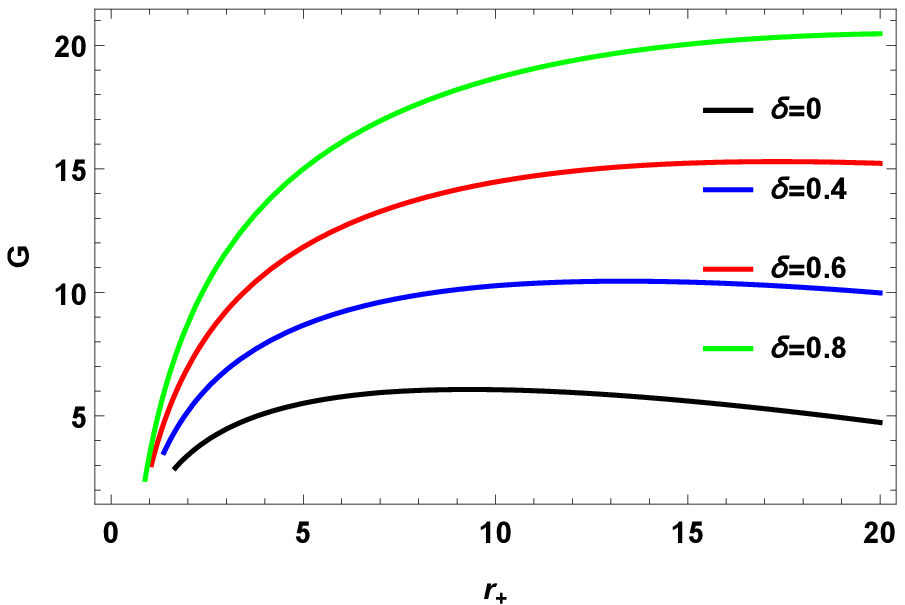}\\
{Figure 7: Gibbs free energy versus $r_{+}$ for a=0.2, Q=0.4.}
\end{center}
The graphical analysis of Gibbs free energy with respect to horizon radius is shows in \textbf{Fig. 7}. The positivity of this energy is sign of occurrence of non-spontaneous reactions means this system requires more energy to gain equilibrium state. After the detail discussion of thermodynamics quantities, there is another important concept is the stability of the system that is checked by specific heat. The specific heat  ($C_{\tilde{S}}=\frac{dE}{dT}$) is given as
\begin{eqnarray}
  C_{\tilde{S}} &=&\Big(2 \Big(a^2+3 r_+^2\Big) \Big(r_+ \Big(r_+ \Big(a^2 \Big(-\delta  (Q-4) \log
   \Big(\frac{\Big(a^2-r_+ \Big(Q+r_+\Big)\Big)^2}{r_+^2 \Big(a^2+r_+^2\Big)}\Big)+a^2 (\pi
   Q-5)+\delta  (Q-4) \log (16 \pi )\Big)+r_{+}\nonumber\\&+& \Big(-\pi  a^4-2 \delta  Q \log \Big(\frac{\Big(a^2-r_+
   \Big(Q+r_+\Big)\Big)^2}{r_+^2 \Big(a^2+r_+^2\Big)}\Big)+r_+ \Big(-\delta  (Q+1) \log
   \Big(\frac{\Big(a^2-r_+ \Big(Q+r_+\Big)\Big)^2}{r_+^2 \Big(a^2+r_+^2\Big)}\Big)+r_+
   \Big(-\delta  \log\nonumber\\&\times& \Big(\frac{\Big(a^2-r_+ \Big(Q+r_+\Big)\Big)^2}{r_+^2
   \Big(a^2+r_+^2\Big)}\Big)+\pi  a^2+\delta  \log (16 \pi )+r_+ \Big(\pi  Q+\pi  r_++1\Big)+2
   Q\Big)+a^2 (2 \pi  Q-3)+\delta  (Q+1) \log (16 \pi )\Big)\nonumber\\&+&2 a^2 Q+2 \delta  Q \log (16 \pi
   )\Big)\Big)-a^4 \Big(-\delta  \log \Big(\frac{\Big(a^2-r_+ \Big(Q+r_+\Big)\Big)^2}{r_+^2
   \Big(a^2+r_+^2\Big)}\Big)+\pi  a^2+\delta  \log (16 \pi )\Big)\Big)-a^4 \Big(-\delta  \log
   \nonumber\\&\times&\Big(\frac{\Big(a^2-r_+ \Big(Q+r_+\Big)\Big)^2}{r_+^2 \Big(a^2+r_+^2\Big)}\Big)+a^2+\delta
   \log (16 \pi )\Big)\Big)\Big)\Big(r_+ \Big(a^2+r_+^2\Big) \Big(-a^4-4 a^2 r_+^2+2 Q r_+^3+r_+^4\Big)\Big)^{-1}.
  \end{eqnarray}
  \begin{center}
  \includegraphics[width=8cm]{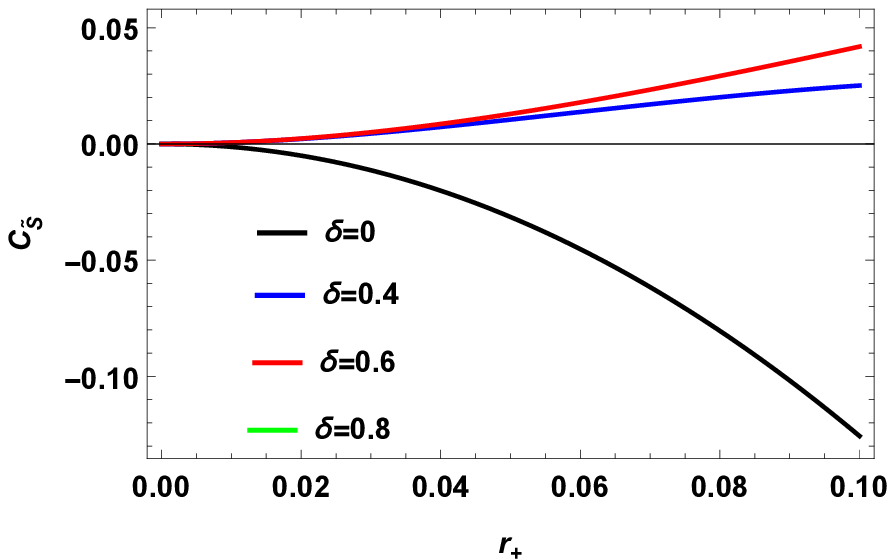}\\
{Figure 8: Specific heat versus $r_{+}$ for a=0.2, Q=0.4.}
\\
\end{center}
 From {\bf Fig. 8}, the behaviour of specific heat is observed with respect to horizon radius and different choices of correction parameter $\delta$. It can be observed that the uncorrected quantity (black) depicts negative behaviour means the system is unstable while the corrected specific heat shows positive behaviour throughout the considered domain. The positivity of this plot is indication of stable region. It can be concluded that these correction terms makes the system stable under thermal fluctuations.

\section{Discussion and Result}

In this paper, we have utilized Lagrangian gravity equation to observe the tunneling of bosonic particles through the horizon of Horndeski like BH. We have used the metric from Ref. \cite{46}. We have considered a new version of black hole with Horndeski parameter $Q$ and a rotation parameter $a$, due to the presence of these parameters, we call the metric a new type of spacetime. Our results are also in terms of these parameters, therefore they are different from the previous literature related about the thermodynamics of this black hole. Assuming to relativistic quantum mechanics and the region of vacuum, where particles are produced continuously in the phenomenon of annihilated. The tunneling radiation as a quantum mechanical processes can be observed as a tunneling phenomenon, where positive boson particle radiate the horizon and the negative energy boson particle move inward and absorbed by the BH. The incoming and outgoing boson particles movement carried out by the action of particle's is real and complex, respectively. The emission rate of these tunneling radiation corresponding to the Horndeski like BH configuration is associated to the imaginary part of the action of particles, which is associated to the factor of Boltzmann, this factor gives $T_{H}$ for Horndeski like BH. From our investigation, we have observed that, in rotating case of BH, the $T_{H}$ at which boson particles tunnel through the BH horizon is not dependent at any types of particles. In special case when particles have different (zero or upward or down) spins, the tunneling rate will be alike by assuming the semi-classical approach. Thus, their corresponding $T_{H}$ must be similar for all types of particle. Therefore, one can say tunneling radiation is
independent of all kinds of the particles and this result also holds for different frame of coordinates by utilizing the transformations of particular coordinate.
For this procedure, the tunneling particles is associated to the energy of particles, momentum, quantum gravity, hairy parameter of Horndeski gravity and BH surface gravity, while the temperature depends on  hairy parameter of Horndeski gravity, rotation and quantum gravity parameters.
It is very important to mention here that, when $\beta=0$, we obtain the standard temperature for Horndeski like BH.
In the absence of charge i.e., $Q=0$, the above temperature reduces into Kerr BH temperature. For $\beta=0$ and $a=0$, the temperature reduces into Reissner Nordstr$\ddot{o}$m BH. Moreover, when $Q=0=a$, we recover the temperature of Schwarzschild BH. For the  changing values of $\beta$ from $10$ to $30$ in the region $0\leq r_{+}\leq5$, we have observed that the Horndeski like BH is stable and for changing $Q$ from $0.5$ to $1.5$, the Horndeski like BH is un-stable with negative temperature. Moreover, the temperature increases with the increasing values of quantum gravity $\beta$.

Moreover, we have computed $T_{H}$ as well as
heat capacity for Horndeski gravity like BHs.
Firstly, the $T_{H}$ is calculated through entropy and the density of state is also calculated with help of inverse Laplace transformation. We have observed that the exact entropy of the system depends on Hawking temperature by applying the method of steepest descent under different conditions. In graphically, we have studied the monotonically increasing entropy of the metric throughout the assumed domain. It is observed that a decrease in entropy for a certain value of the radius, the corrected expression of energy is increasing smoothly and also studied that the small BHs are more effective for thermal fluctuations.

It is observed that the behaviour of energy gradually decreases for the different correction parameter $\delta$ values and the graph of usual entropy shows opposite behaviour as the graph increases. Therefore, the considered system shifts its state towards equilibrium.
It is observable that for the small values of radii, the graph gradually decreases even shifts towards negative side, while the corrected internal energy depicts positive behaviour. So, the considered BH absorbed more and more heat from the surrounding to maintain its state.
The graph of pressure coincides the state of equilibrium. For the different values of correction parameter, the pressure significantly increases for the considered system.

It is observed that the graph of usual enthalpy coincides with the plots of corrected one and abruptly decreases even shifts towards negative side. It can be concluded that there exists a exothermic reactions means a huge amount of energy released into its surroundings.
The positivity of this energy is sign of occurrence of non-spontaneous reactions means the system required more energy to gain equilibrium state.
The behaviour of specific heat with respect to horizon radius and different choices of correction parameter $\delta$ has been observed. The uncorrected quantity depicts negative behaviour means the system is unstable while the corrected specific heat shows positive behaviour throughout the considered domain. The positivity of the plots for specific heat is indication of stable region. It can be concluded that these correction terms makes the system stable under thermal fluctuations.

\section*{Appendix A}
We have utilized the Lagrangian equation in the approximation of WKB to get following solutions,

\begin{eqnarray}
&&\frac{I}{(R^2+fI)g^{-1}}
\left[\eta_{1}(\partial_{0}K_{0})(\partial_{1}K_{0})+\beta \eta_{1}
(\partial_{0}K_{0})^{3}(\partial_{1}K_{0})-\eta_{0}(\partial_{1}K_{0})^{2}
-\beta \eta_{0}(\partial_{1}K_{0})^4+\eta_{1}eA_{0}(\partial_{1}K_{0})\right.+\nonumber\\
&&\left.\eta_{1}\beta eA_{0}(\partial_{0}K_{0})^{2}(\partial_{1}K_{0})\right]
-\frac{R}{g^{-1}(fI+R^2)}
\left[\eta_{3}(\partial_{1}K_{0})^2+\beta \eta_{3}(\partial_{1}K_{0})^4
-\eta_{1}(\partial_{1}K_{0})(\partial_{3}K_{0})-\beta \eta_{1}
(\partial_{1}K_{0})(\partial_{3}K_{0})^2\right]\nonumber\\
&&+\frac{I}{h(fI+R^2)}
\left[\eta_{2}(\partial_{0}K_{0})(\partial_{2}K_{0})+\beta \eta_{2}
(\partial_{0}K_{0})^3(\partial_{2}K_{0})-\eta_{0}(\partial_{2}K_{0})^2
-\beta \eta_{0}(\partial_{2}K_{0})^4+\eta_{2}eA_{0}(\partial_{2}K_{0})\right.\nonumber\\
&&+\left.\eta_{2}eA_{0}\beta(\partial_{0}K_{0})^{2}(\partial_{1}K_{0})\right]
+\frac{fI}{(fI+R^2)^2}
\left[\eta_{3}(\partial_{0}K_{0})(\partial_{3}K_{0})+\beta \eta_{3}
(\partial_{0}K_{0})^{3}(\partial_{3}K_{0})-\eta_{0}(\partial_{3}K_{0})^{2}
-\beta \eta_{0}(\partial_{3}K_{0})^4\right.\nonumber\\
&&+\left. \eta_{3}eA_{0}(\partial_{3}K_{0})+\eta_{3}eA_{0}(\partial_{0}K_{0})^{2}
(\partial_{3}K_{0})\right]-m^2\frac{I \eta_{0}-R \eta_{3}}
{(fI+R^2)}=0,\label{j1}\\
&&\frac{-I}{g^{-1}(fI+R^2)}
\left[\eta_{1}(\partial_{0}K_{0})^2+\beta \eta_{1}(\partial_{0}K_{0})^4-\eta_{0}
(\partial_{0}K_{0})(\partial_{1}K_{0})-\beta \eta_{0}(\partial_{0}K_{0})(\partial_{1}K_{0})^{3}
+\eta_{1}eA_{0}(\partial_{0}K_{0})
+\beta \eta_{1}eA_{0}(\partial_{0}K_{0})^3\right]\nonumber\\
&&+\frac{R}{g^{-1}
(fI+R^2)}\left[\eta_{3}(\partial_{0}K_{0})(\partial_{1}K_{0})
+\beta \eta_{3}(\partial_{0}K_{0})
(\partial_{1}K_{0})^3-\eta_{1}(\partial_{0}K_{0})(\partial_{3}K_{0})
-\beta \eta_{1}(\partial_{0}K_{0})(\partial_{3}K_{0})^{3}\right]\nonumber\\
&&+\frac{1}{g^{-1}h}\left[\eta_{2}(\partial_{1}K_{0})(\partial_{2}K_{0})\right.
+\left.\beta \eta_{2}
(\partial_{1}K_{0})(\partial_{2}K_{0})^3-\eta_{1}(\partial_{2}K_{0})^{2}
-\beta \eta_{1}(\partial_{2}K_{0})^{4}\right]\nonumber\\&&+\frac{1}{g^{-1}(fI
+R^2)}\left[\eta_{3}(\partial_{1}K_{0})(\partial_{3}K_{0})
+\beta \eta_{3}(\partial_{1}K_{0})(\partial_{3}K_{0})^3-\eta_{1}
(\partial_{3}K_{0})^2\right.-\left.\beta \eta_{1} (\partial_{3}K_{0})^{4}\right]-\frac{m^2 \eta_{1}}{g^{-1}}
+\frac{eA_{0}I}{g^{-1}(fI+R^2)}\nonumber\\&&
\left[\eta_{1}(\partial_{0}K_{0})+\beta \eta_{1}(\partial_{0}K_{0})^3
-\eta_{0}(\partial_{1}K_{0})-\beta \eta_{0}(\partial_{1}K_{0})^3
+ eA_{0}\eta_{1}\right.
+\left.\beta \eta_{1}eA_{0}(\partial_{0}K_{0})^{2})\right]\nonumber\\
&&+\frac{eA_{0}R}{g^{-1}(fI+R^2)}
\left[\eta_{3}(\partial_{1}K_{0})+\beta \eta_{3}(\partial_{1}K_{0})^3
-\eta_{1}(\partial_{3}K_{0})-\beta \eta_{1}(\partial_{1}K_{0})^3\right]=0,
\end{eqnarray}
\begin{eqnarray}
&&\frac{I}{h(fI+R^2)}
\left[\eta_{2}(\partial_{0}K_{0})^2+\beta \eta_{2}(\partial_{0}K_{0})^{4}
-\eta_{0}(\partial_{0}K_{0})(\partial_{2}K_{0})-\beta \eta_{0}(\partial_{0}K_{0})(\partial_{2}K_{0})^3
+\eta_{2}eA_{0}(\partial_{0}K_{0})+\beta \eta_{2}eA_{0}(\partial_{0}K_{0})^{3}\right]\nonumber\\
&&+\frac{1}{g^{-1}h}\left[\eta_{2}(\partial_{1}K_{0})^2+\beta \eta_{2}
(\partial_{1}K_{0})^{4}-\eta_{1}(\partial_{1}K_{0})(\partial_{2}K_{0})
-\beta \eta_{1}(\partial_{1}K_{0})(\partial_{2}K_{0})^3\right]-\frac{R}
{h(fI+R^2)}\Big[\eta_{2}(\partial_{0}K_{0})
(\partial_{3}K_{0})\nonumber\\
&&+\left.\beta \eta_{2}(\partial_{0}K_{0})^{3}(\partial_{3}K_{0})-\eta_{0}
(\partial_{0}K_{0})(\partial_{3}K_{0})-\beta \eta_{0}(\partial_{0}K_{0})^3
(\partial_{3}K_{0})+\eta_{2}eA_{0}(\partial_{3}K_{0})+\beta \eta_{2}eA_{0}
(\partial_{3}K_{0})^{3}\right]\nonumber\\
&&+\frac{f}{h(fI+R^2)}\left[\eta_{3}
(\partial_{2}K_{0})(\partial_{3}K_{0})+\beta \eta_{3}
(\partial_{2}K_{0})^{3}(\partial_{3}K_{0})-\eta_{2}(\partial_{3}K_{0})^2
-\beta \eta_{2}(\partial_{3}K_{0})^4\right]+\frac{eA_{0}I}{h
(fI+R^2)}\Big[\eta_{2}(\partial_{0}K_{0})\nonumber\\
&&+\left.\beta \eta_{2}(\partial_{0}K_{0})^3-\eta_{0}(\partial_{2}K_{0})
-\beta \eta_{0}(\partial_{2}K_{0})^3+\eta_{2}eA_{0}+\eta_{2}\beta eA_{0}
(\partial_{0}K_{0})^2\right]-\frac{m^2 \eta_{2}}{h}=0,\\
&&\frac{(fI-f^2)}{(fI
+R^2)^2}\left[\eta_{3}(\partial_{0}K_{0})^2+\beta \eta_{3}
(\partial_{0}K_{0})^4-\eta_{0}(\partial_{0}K_{0})(\partial_{3}K_{0})
-\beta \eta_{0}(\partial_{0}K_{0})(\partial_{3}K_{0})^{3}+{eA_{0}\eta_3}
(\partial_{0}K_{0})+\beta \eta_{3}eA_{0}(\partial_{0}K_{0})^{3}\right]\nonumber\\
&&-\frac{I}{h(fI+R^2)}
\left[\eta_{3}(\partial_{1}K_{0})^2+\beta \eta_{3}(\partial_{1}K_{0})^{4}
-\eta_{1}(\partial_{1}K_{0})(\partial_{3}K_{0})-\beta \eta_{1}(\partial_{1}
K_{0})(\partial_{3}K_{0})^3\right]-\frac{R}{h(f
I+R^2)}\nonumber\\&&\Big[\eta_{2}(\partial_{0}K_{0})(\partial_{2}K_{0})
+\left.\beta \eta_{2}(\partial_{0}K_{0})^3(\partial_{2}K_{0})-\eta_{0}
(\partial_{2}K_{0})^{2}+\beta \eta_{0}(\partial_{2}K_{0})^4+{eA_{0}\eta_2}
(\partial_{2}K_{0})+\beta \eta_{2}eA_{0}\partial_{0}K_{0})^{2}(\partial_{2}
K_{0})\right]\nonumber\\&&-\frac{eA_{0}f}{h(fI+R^2)}
\left[\eta_{3}(\partial_{2}K_{0})^2+\beta \eta_{3}(\partial_{2}K_{0})^4-\eta_{2}(\partial_{2}K_{0})(\partial_{3}K_{0})
-\beta \eta_{2}(\partial_{0}K_{0})(\partial_{3}K_{0})^{3}\right]
-\frac{m^2 (R \eta_{0}-f\eta_{3}}{(fI+R^2)}\nonumber\\&&
+\frac{eA_{0}(fI-f^2)}{(fI+R^2)^2}
\left[\eta_{3}(\partial_{0}K_{0})+\beta \eta_{3}
(\partial_{0}K_{0})^3
-\eta_{0}(\partial_{3}K_{0})-\beta \eta_{0}(\partial_{3}K_{0})^3+\eta_{3}eA_{0}
+\eta_{3}\beta eA_{0}(\partial_{0}K_{0})^2\right]=0,\label{j2}
\end{eqnarray}

\section*{Appendix B}

The thermodynamic quantity enthalpy is given as

\begin{eqnarray}
H&=&\Big(r_+ \Big(a^2+r_+^2\Big) \Big(r_+ \Big(a^2+r_+^2\Big) \Big(r_+ \Big(Q+r_+\Big)-a^2\Big)\Big(\delta  \Big(\log (16 \pi )-\log \Big(\frac{\Big(a^2-r_+ \Big(Q+r_+\Big)\Big)^2}{r_+^2
\Big(a^2+r_+^2\Big)}\Big)\Big)\nonumber\\&+&\pi  \Big(a^2+r_+^2\Big)\Big)-\Big(a^4-r_+^2 \Big(-4 a^2+2 Q r_++r_+^2\Big)\Big) \Big(-\delta  \log \Big(\frac{\Big(a^2-r_+
   \Big(Q+r_+\Big)\Big)^2}{r_+^2 \Big(a^2+r_+^2\Big)}\Big)+a^2+\delta  \log (16 \pi
   )+r_+^2\Big)\Big)\nonumber\\&+&\Big(2 r_+^2 \Big(a^2+r_+^2\Big) \Big(-4 a^2+3 Q r_++2 r_+^2\Big)
   \Big(a^2-r_+ \Big(Q+r_+\Big)\Big)^2 \Big(-\delta  \log \Big(\frac{\Big(a^2-r_+
   \Big(Q+r_+\Big)\Big)^2}{r_+^2 \Big(a^2+r_+^2\Big)}\Big)+a^2\nonumber\\&+&\delta  \log (16 \pi
   )+r_+^2\Big)-2 \Big(r_+ \Big(Q+r_+\Big)-a^2\Big) \Big(a^4-r_+^2 \Big(-4 a^2+2 Q
   r_++r_+^2\Big)\Big) \Big(-a^4 \delta +Q r_+^3 \Big(a^2+\delta \Big)\nonumber\\&-&r_+^2 \Big(a^4+3 a^2 \delta
   \Big)+Q r_+^5+r_+^6\Big)+4 r_+^2 \Big(a^4-r_+^2 \Big(-4 a^2+2 Q r_++r_+^2\Big)\Big)
   \Big(a^2-r_+ \Big(Q+r_+\Big)\Big)^2 \Big(-\delta  \log \nonumber\\&\times&\Big(\frac{\Big(a^2-r_+
   \Big(Q+r_+\Big)\Big){}^2}{r_+^2 \Big(a^2+r_+^2\Big)}\Big)+a^2+\delta  \log (16 \pi
   )+r_+^2\Big)+2 \Big(a^2+r_+^2\Big) \Big(a^4-r_+^2 \Big(-4 a^2+2 Q r_++r_+^2\Big)\Big)\nonumber\\&\times&
   \Big(a^2-r_+ \Big(Q+r_+\Big)\Big)^2 \Big(-\delta  \log \Big(\frac{\Big(a^2-r_+
   \Big(Q+r_+\Big)\Big)^2}{r_+^2 \Big(a^2+r_+^2\Big)}\Big)+a^2+\delta  \log (16 \pi
   )+r_{+}^{2}\Big)\Big)\Big(\Big(a^{2}-r_{+} \Big(Q+r_+\Big)\Big)^2\Big)^{-1})\nonumber\\&\times&\Big(4 \pi  r_+^3 \Big(a^2+r_+^2\Big)^3\Big)^{-1}.
  \end{eqnarray}
The Gibbs free energy is expressed as \begin{eqnarray}
  G&=&\Big(-r_+^2 \Big(a^2+r_+^2\Big){}^2 \Big(r_+ \Big(Q+r_+\Big)-a^2\Big) \Big(-\delta  \log
   \Big(\frac{\Big(a^2-r_+ \Big(Q+r_+\Big)\Big){}^2}{r_+^2 \Big(a^2+r_+^2\Big)}\Big)+a^2+\delta
   \log (16 \pi )+r_+^2\Big)+r_+ \Big(a^2+r_+^2\Big)\nonumber\\&\times& \Big(r_+ \Big(a^2+r_+^2\Big) \Big(r_+
   \Big(Q+r_+\Big)-a^2\Big) \Big(\delta  \Big(\log (16 \pi )-\log \Big(\frac{\Big(a^2-r_+
   \Big(Q+r_+\Big)\Big){}^2}{r_+^2 \Big(a^2+r_+^2\Big)}\Big)\Big)+\pi
   \Big(a^2+r_+^2\Big)\Big)-\Big(a^4-r_+^2 \nonumber\\&\times& \Big(-4 a^2+2 Q r_++r_+^2\Big)\Big) \Big(-\delta
   \log \Big(\frac{\Big(a^2-r_+ \Big(Q+r_+\Big)\Big){}^2}{r_+^2
   \Big(a^2+r_+^2\Big)}\Big)+a^2+\delta  \log (16 \pi )+r_+^2\Big)\Big)+\Big(2 r_+^2
   \Big(a^2+r_+^2\Big) \nonumber\\&\times& \Big(-4 a^2+3 Q r_++2 r_+^2\Big) \Big(a^2-r_+ \Big(Q+r_+\Big)\Big){}^2
   \Big(-\delta  \log \Big(\frac{\Big(a^2-r_+ \Big(Q+r_+\Big)\Big){}^2}{r_+^2
   \Big(a^2+r_+^2\Big)}\Big)+a^2+\delta  \log (16 \pi )+r_+^2\Big)\nonumber\\&-&2 \Big(r_+
   \Big(Q+r_+\Big)-a^2\Big) \Big(a^4-r_+^2 \Big(-4 a^2+2 Q r_++r_+^2\Big)\Big) \Big(-a^4 \delta
   +Q r_+^3 \Big(a^2+\delta \Big)-r_+^2 \Big(a^4+3 a^2 \delta \Big)+Q r_+^5+r_+^6\Big)\nonumber\\&+&4 r_+^2
   \Big(a^4-r_+^2 \Big(-4 a^2+2 Q r_++r_+^2\Big)\Big) \Big(a^2-r_+ \Big(Q+r_+\Big)\Big)^2
   \Big(-\delta  \log \Big(\frac{\Big(a^2-r_+ \Big(Q+r_+\Big)\Big){}^2}{r_+^2
   \Big(a^2+r_+^2\Big)}\Big)+a^2+\delta  \log (16 \pi )\nonumber\\&+& r_+^2\Big)+2 \Big(a^2+r_+^2\Big)
   \Big(a^4-r_+^2 \Big(-4 a^2+2 Q r_++r_+^2\Big)\Big) \Big(a^2-r_+ \Big(Q+r_+\Big)\Big)^2
   \Big(-\delta  \log \Big(\frac{\Big(a^2-r_+ \Big(Q+r_+\Big)\Big){}^2}{r_+^2
   \Big(a^2+r_+^2\Big)}\Big)\nonumber\\&+&a^2+\delta  \log (16 \pi )+r_+^2\Big)\Big)\Big(\Big(a^2-r_+
   \Big(Q+r_+\Big)\Big){}^2\Big)\Big)\Big(4 \pi  r_+^3 \Big(a^2+r_+^2\Big){}^3\Big)^{-1}.
   \end{eqnarray}

\end{document}